\documentclass[12pt,twocolumn]{emulateapj}
\usepackage{apjfonts}
\def\lsim{\lower 2pt \hbox{$\, \buildrel {\scriptstyle <}\over
{\scriptstyle \sim}\,$}}
\def\gsim{\lower 2pt \hbox{$\, \buildrel {\scriptstyle >}\over
{\scriptstyle \sim}\,$}}

\slugcomment{Accepted for publication in ApJ}

\shorttitle{MHD Accretion Flows: Magnetic Tower Jet and
  Quasi-Steady State}
\shortauthors{Kato, Mineshige, \& Shibata}

\begin{document}

\title{Magnetohydrodynamical Accretion Flows:
       Formation of Magnetic Tower Jet and Subsequent Quasi-Steady
       State}

\author{Y. kato, S. Mineshige} 
\affil{Yukawa Institute for Theoretical Physics, Kyoto University,
Kyoto 606-8502, Japan}
\and
\author{K. Shibata}
\affil{Kwasan Observatory, Kyoto University, Yamashina-ku, Kyoto
    607-8471, Japan}

\begin{abstract}
We present three-dimensional (3-D) magnetohydrodynamical (MHD)
simulations of radiatively inefficient accretion flow around black
holes.  General relativistic effects are simulated by using the
pseudo-Newtonian potential.  We start calculations with a rotating
torus threaded by localized poloidal magnetic fields with
plasma beta, a ratio of the gas pressure to the magnetic pressure,
$\beta =10$ and $100$.  When the bulk of torus material reaches the
innermost region close to a central black hole, a magnetically driven
jet emerges.  This magnetic jet is derived by vertically inflating
toroidal fields (`magnetic tower') and has a two-component structure:
low-$\beta$ ($\lsim 1$) plasmas threaded with poloidal (vertical)
fields are surrounded by that with toroidal fields.  The collimation
width of the jet depends on external pressure, pressure of ambient
medium; the weaker the external pressure is, the wider and the
longer-lasting becomes the jet.  Unless the external pressure is
negligible, the bipolar jet phase ceases after several dynamical
timescales at the original torus position and a subsequent
quasi-steady state starts.  The black hole is surrounded by
quasi-spherical zone with highly inhomogeneous structure in which
toroidal fields are dominant except near the rotation axis.  Mass
accretion takes place mainly along the equatorial plane. Comparisons
with other MHD simulation results and observational implications are
discussed.
\end{abstract}


\section{INTRODUCTION}
The study of accretion disk structure has a long research history.
Yet, fundamental structure has not become clear.  The standard disk
model established in the early 1970's has been very successful in
describing the high/soft states of Galactic black hole (BH) candidates
(see, e.g., Ebisawa 1999), but their low/hard state characterized by
power-law type spectra was problematic to the standard model.
Alternative disk models which account for high-energy emissions from
X-ray binaries (XRBs) and active galactic nuclei (AGNs)
have been extensively discussed and the base of a distinct type of disk
models has been constructed by Narayan \& Yi (1994), which turned out
to be the same model as that proposed by Ichimaru (1977).  This notion
is now referred to as ``Advection Dominated Accretion Flow'' or
``ADAF'' (see reviews by Narayan, Mahadevan, \& Quataert 1998; Kato,
Fukue, \& Mineshige 1998).

The key relation to discriminate the standard and ADAF solutions
resides in the energy equation of accretion disks, $Q_{\rm adv} =
Q^-_{\rm vis} - Q^+_{\rm rad}$; from the left, the advection term
representing energy transport carried by accreting material, viscous
heating, radiative cooling terms, respectively.  This relation yields
three distinct solutions, depending on the final forms of accretion
energy.  Accretion energy goes to radiation in a standard disk so that
the disk becomes cool and shine bright.  In an optically thin ADAF, in
contrast, accretion energy is converted to internal energy of gas.
ADAF is thus characterized by high temperature and low emissivity.
The third is a hot optically thin cooling-dominated disk
  (Shapiro, Lightman, \& Eardley 1976) which is thermally unstable.

Although the ADAF model is quite successful in reproducing the hard
spectra of Galactic BH candidates, as well as those of low-luminosity
AGNs and our Galactic center source, Sgr A*, it comprises a serious
problem.  The original optically thin ADAF model has been constructed
basically in a vertically one-zone approximation. In other words, it
was formulated in one (radial) dimension, although multi-dimensional
flow patterns seem to be essential.  In this regard, Narayan and Yi
(1994) already made two important pieces of predictions: possible
occurrence of convection (in the radial direction) and outflows.
Since radiative cooling is inefficient, entropy of accreting gas
should monotonically increase towards a black hole (i.e., in the
direction of gravity), a condition for a convection.  Further, the
self-similar model points an advection-dominated flow having positive
Bernoulli parameter, $Be$, meaning that matter is gravitationally
unbound and can form outflows.

Actually, 2- or 3-D hydrodynamical simulations exhibit complex flow
  patterns (Stone, Pringle, \& Begelman 1999; see also, Igumenshchev
  \& Abramowicz 2000, Igumenshchev, Abramowicz, \& Narayan
  2000). Taking this fact into account, several authors proposed
  alternative models with some modifications to the ADAF picture.
  Blandford \& Begelman (1999), for example, constructed a model of
  ADIOS (Adiabatic Inflow/Outflow Solution) so as to incorporate the
  emergence of strong jets in the one-dimensional scheme.  CDAF
  (Quataert \& Gruzinov 2000; Narayan 2002), on the other hand,
  considers the significant effects of large-scale convective motion.
  Especially, they point out that angular momentum is carried not
  outward (as in the viscous disk) but inward, while energy is
  transported outward in CDAF.  These are the main products in the
  second stage of research of hot accretion flow (or radiatively
  inefficient flow).

Yet, this is not the end of the story.  One may well ask what an ADAF
model can properly treat since it is a time-stationary, 1D, alpha
model.  Note that the ADIOS and CDAF models also share that problem:
they also cannot properly treat the accretion flows, unlike a full
numerical simulation.

Furthermore, hydrodynamical approach seems to be totally inappropriate
in radiatively inefficient flow.  Rather, we expect that magnetic
fields play crucial roles in hot accretion flows; disks in low/hard
state could be low-$\beta$ disk corona, in which magnetic pressure
exceeds gas pressure (Mineshige, Kusunose, \& Matsumoto 1995).  That
is, magnetohydrodynamical (MHD) approach is indispensable.  [To be
  more precise, the significance of magnetic fields is not restricted
  to radiatively inefficient flow but also in a standard-type disk,
  since magnetic fields provide source of disk viscosity as a
  consequence of magneto-rotational instability (MRI; Balbus \& Hawley
  1991).]  Now we have entered the third stage of the research of hot
accretion flow based on the multi-dimensional MHD simulations.

First global 3-D MHD simulations of accretion disks have been made by
Matsumoto (1999).  He calculated the evolution of magnetic fields and
structural changes of a torus which is initially threaded by toroidal
magnetic fields.  In relation to the MHD accretion flow, we would like
to remind readers that similar simulations had already been performed
by many groups in the context of astrophysical jets, pioneered by
Uchida \& Shibata (1985; see also Shibata \& Uchida 1986).  They
calculated the evolution of a disk threaded with vertical fields
extending to infinity  to see how a magnetic jet is produced.  Hence,
their simulations were more concerned with the jet structure and a
disk only plays a passive role, but it is instructive to see how the
disk behaves in those simulations.  Since angular momentum can be very
efficiently extracted from the surface of the accretion disks by the
vertical fields, a surface avalanche produces anomalous mass accretion
in those simulations.  Thus, we need to be careful that the situation
may depend on whether magnetic fields are provided externally or they
are generated internally.

Many MHD disk simulations starting with locally confined fields have
been published after 2000.  Machida, Matsumoto, \& Hayashi (2000)
extended Matsumoto (1999) and studied the structure of MHD flow
starting with initially toroidal, localized fields and found global
low-$\beta$ filaments created in the turbulent accretion disk 
(see also Hawley 2000; Hawley \& Krolik 2001).  Machida, Matsumoto,
and Mineshige (2001) have shown that its flow structure is very
similar to those of CDAF in the sense of visible convective patterns
and relatively flatter density profile ($\rho\propto r^{-1/2}$), which
contrasts with $\rho \propto r^{-3/2}$ in ADAF.  They conjectured
that such flow pattern is driven by thermodynamic buoyant
instabilitiesis, however, Stone \& Pringle (2001) established that it
is most likely to be produced by the turbulence due to the MRI (see
also Hawley, Balbus, \& Stone 2001; Hawley 2001).  Thus, the MRI is
the fundermental process that converts gravitational energy into
turbulent energy within the magnetized disk (see Balbus 2003 for a
review).

Outflows also appear in MHD simulations starting with localized fields.
Kudoh, Matsumoto, \& Shibata (2002; hereafter refered to as KMS02),
for example, have found the rising magnetic loop which behaves like a
jet (see also Turner, Bodenheimer, \& R{\' o}{\.z}yczka 1999).  They
assert that the jet is collimated by a pinching force of the toroidal
magnetic field, and that its velocity is on order of the Keplerian
velocity of the disk.  Hawley \& Balbus (2002; hereafter referred to
as HB02) calculated the evolution of a torus with initial poloidal
fields and found three well-defined dynamical components: a hot,
thick, rotationally dominated, high-$\beta$ Keplerian disk; a
surrounding hot, unconfined, low-$\beta$ coronal envelope; and a
magnetically confined unbound high-$\beta$ jet along the centrifugal
funnel wall (see also Kudoh, Matsumoto, Shibata 1998; Stone \& Pringle
2001; Hawley, Balbus, \& Stone 2001; Casse \& Keppens 2002).

Recently, Igumenshchev, Narayan \& Abramowicz (2003; hereafter
referred to as INA03), have reported that the flow structure with
continuous mass and toroidal or poloidal field injection evolves
through two distinct phases: an initial transient phase associated
with a hot corona and a bipolar outflow, and a subsequent steady
state, with most of the volume being dominated by a strong dipolar
magnetic field.  In addition, they also have argued that the accretion
flow is totally inhibited by the strong magnetic field.

In the present study, we performed 3-D MHD simulations of a rotating
  torus with the same initial condition as that of HB02, but adopting
  the different inner boundary condition as that of HB02.  The aims of
  the present study are two-fold: the first one is to examine how an
  MHD jet emerges from localized field configurations and what
  properties it has.  The second one is to elucidate the dynamics of
  MHD accretion flow in radiation-inefficient regimes by means of
  long-term simulations and compare our results with previous ones.
  In \S 2 we present basic equations and explain our models of 3-D
  global MHD simulation.  We then present our results in the first
  phase of an evolving magnetic-tower jet and in the second phase of a
  subsequent quasi-steady  in \S 3.  The final section is devoted to
  brief summary and discussion.

\section{OUR MODELS AND METHODS OF SIMULATIONS}
We solve the basic equations of the resistive magnetohydrodynamics
in the cylindrical coordinates, $(r,\phi,z)$.  General relativistic
effects are incorporated by the pseudo-Newtonian potential
(Paczy\'{n}sky \& Wiita 1980), $\psi=-GM/(R-r_{\rm s})$, where
$R$($\equiv \sqrt{r^{2}+z^{2}}$) is the distance from the origin, and
$r_{\rm s}$ $(\equiv 2GM/c^{2})$ is the Schwarzschild radius (with $M$
and $c$ being the mass of a black hole and the speed of light,
respectively).  The basic equations are then written in a conservative
form as follows: 
\begin{equation}
  {\partial\rho\over \partial t}+\mbox{\boldmath$\nabla$}\cdot(\rho
  \mbox{\boldmath$v$}) = 0,
\label{continuity}
\end{equation}
\begin{eqnarray}
  {\partial\over\partial t}\left(\rho\mbox{\boldmath $v$}\right)
  +\mbox{\boldmath$\nabla$}\cdot\left(\rho
  \mbox{\boldmath$v$}\mbox{\boldmath$v$} -
  {\mbox{\boldmath$B$}\mbox{\boldmath$B$}\over 4\pi}\right)
  =-\mbox{\boldmath$\nabla$}\left(p_{\rm gas} +
  \frac{B^2}{8\pi}\right)-\rho\mbox{\boldmath$\nabla$}\psi,
\label{momentum}
\end{eqnarray}
\begin{eqnarray}
  {\partial\over\partial t}\left(\epsilon +{B^{2}\over 8\pi}\right)
    +\mbox{\boldmath$\nabla$}\cdot\left[\left(\epsilon+p_{\rm gas}
    \right)\mbox{\boldmath$v$}+{\mbox{\boldmath$E$}\times
    \mbox{\boldmath$B$}\over 4\pi}\right]
  =-\rho\mbox{\boldmath$v$}\cdot\mbox{\boldmath$\nabla$}\psi,
\label{energy}
\end{eqnarray}
and
\begin{equation}
{\partial\mbox{\boldmath$B$}\over \partial t} =
- c \mbox{\boldmath$\nabla$} \times \mbox{\boldmath$E$}
\label{induction}
\end{equation}
where $\epsilon=\rho v^{2}/2+p_{\rm gas}/(\gamma-1)$ 
is the energy of the gas, and $\mbox{\boldmath$E$} =
-(\mbox{\boldmath$v$}/c)\times \mbox{\boldmath$B$} +
(4\pi\eta/c^{2})\mbox{\boldmath$J$}$ is the Ohm's law.
$\mbox{\boldmath$J$} = (c/4\pi)\mbox{\boldmath$\nabla$} \times
\mbox{\boldmath$B$}$ is electric current.  We fix the adiabatic index
to be $\gamma=5/3$. As to the resistivity, we assign the anomalous
resistivity, which is used in many solar flare simulations (e.g.,
Yokoyama \& Shibata 1994):
\begin{equation}
  \eta = \left\{ \begin{array}{lcl}
    0  & \mbox{for} & v_{\rm d} < v_{\rm crit} \\   
   \eta_{\rm max} [(v_{\rm d}/v_{\rm crit})-1]^2             
   & \mbox{for} & v_{\rm crit} < v_{\rm d} <2v_{\rm crit} \\  
   \eta_{\rm max} & \mbox{for} & v_{\rm d} \geq 2 v_{\rm crit} \\ 
    \end{array} \right. 
\end{equation}
where $v_{\rm d} \equiv |\mbox{\boldmath$J$}|/\rho$ is electron drift
velocity, $v_{\rm crit}$ is the critical velocity, over which
anomalous resistivity turns on, and $\eta_{\rm max}$ is the maximum
resistivity.  In the present study, we assign $v_{\rm crit} = 0.01$
(in the unit of $c$) and $\eta_{\rm max}=10^{-3} c~r_{\rm s}$;
i.e., magnetic Reynolds number is $R_{\rm m}\approx c r_{\rm
  s}/\eta_{\rm max} = 10^3$ in the diffusion region.  Note that the
critical resistivity due to numerical diffusion in our code is much
less, corresponding to $\eta_{\rm c}=10^{-5} c~r_{\rm s}$.  Hence, we
find $\eta_{\rm max}=100\eta_{\rm c}$.  The entropy of the gas can
increase due to the dissipation of the magnetic energy and also the
adiabatic compression in the shock.

We start calculations with a rotating torus in hydrostatic balance,
which was calculated based on the assumption of a polytropic equation
of state, $p=K\rho^{1+1/n}$ with $n=3$, and a power-law specific
angular momentum distribution, $l=l_{0}(r/r_{0})^{a}$ with
$l_{0}=\sqrt{GM r_{0}^{3}}/(r_{0}-r_{\rm s})$.  Here, $a$ and $r_0$
being parameters (assigned later).  Then, the initial density and
pressure distributions of the torus are explicitly written as
\begin{equation}
  \rho_{t}(r,\phi,z)=\rho_{0}\left(1-{\gamma\over v^{2}_{\rm s,0}}
               {\tilde{\psi}-\tilde{\psi}_{0}\over n+1}\right)^{n},
\end{equation}
and
\begin{equation}
  p_{\rm t}(r,\phi,z)=\rho_{0}\frac{v_{\rm s,0}^2}{\gamma}
  \left[\frac{\rho_{\rm t}(r,\phi,z)}{\rho_{0}}\right]^{1+1/n},
\end{equation}
where $\rho_{0}$, $v_{\rm s,0}$, and $\tilde{\psi}$ denote,
respectively, the initial density at $(r,z) = (r_0,0)$, sound velocity
in the torus, and the effective potential given by
\begin{equation}
   \tilde{\psi}(r,z)=\psi(r,z)+\frac{1}{2(1-a)}
   \left(\frac{l}{r}\right)^{2}.
\end{equation}
In the present study, we assign $a = 0.46$ and $r_0=40$ (in the unit
of $r_{s}$) where the Keplerian orbital time is $1124 r_{s}/c$.

Outside the torus, we assume non-rotating, spherical, and isothermal
hot background (referred to as a background corona to distinguish
from a disk corona created at later times as a result of magnetic
activity within a flow) that is initially in hydrostatic equilibrium.
This background corona asserts external pressure to magnetic jets.
The density and pressure distribution of the background corona are:
\begin{equation}
  \rho_{\rm c}(r,\phi,z)
     =\rho_1\exp\left[-\frac{\psi(r,z)-\psi_1}{v^{2}_{\rm
     s,c}/\gamma}\right],
\end{equation}
and
\begin{equation}
  p_{\rm c}(r,\phi,z)=\rho_{\rm c}(r,\phi,z)v^{2}_{\rm s,c}/\gamma,
\label{Pcorona}
\end{equation}
respectively,
where $\rho_{\rm 1}$ and $\psi_1$ are density and potential at
$(r,z)=(r_1,0)$ with $r_1=2 r_{\rm s}$ being the innermost radius,
respectively, while $v_{\rm s,c}$ is (constant) sound speed in the
corona.

The initial magnetic fields are confined within a torus and purely
poloidal.  The initial field distribution is described in terms of the
$\phi$-component of the vector potential which is assumed to be
proportional to the density; that is $A_{\phi} \propto\rho$. (All
other components are zero.)  The strength of the magnetic field is
represented by the plasma-$\beta$, ratio of gas pressure to magnetic
pressure, which is constant in the initial torus.  Initial conditions
are characterized by several non-dimensional parameters, which are
summarized in Tables 1 and 2.  Here, $E_{\rm th,0} \equiv v^{2}_{\rm
  s, 0}/\gamma|\psi_{0}|$ and $E_{\rm th,c}\equiv v^{2}_{\rm s,
  c}/\gamma|\psi_{1}|$, respectively, represent the ratios of thermal
energy to gravitational energy in the initial torus and in the corona.

\begin{deluxetable}{cll}
\tablecolumns{3}
\tablewidth{0pc}
\tablecaption{Basic parameters.}
\tablehead{
\colhead{Parameter}
& \colhead{Definition} 
& \colhead{Adopted values}
}
\startdata
 plasma-$\beta$   & gas pressure/magnetic pressure  & 10 or 100  \\
  $\rho_1/\rho_0$  
& density at $(r_1,0)$/density at $(r_0,0)$ 
& $2.0 \times 10^{-5}$ \\
 $E_{\rm th,0}$ 
&  $\equiv v^2_{\rm s,0}/\gamma|\psi_0|$
& $1.45 \times 10^{-3}$ \\
 $E_{\rm th,c}$ 
&  $\equiv v^2_{\rm s,c}/\gamma|\psi_1|$ 
& 1.0 or 0.1  \\
 $v_{\rm s,0}/c$ 
& sound velocity at $(r_0,0)$ 
& 5.6 $\times 10^{-3}$ \\
 $v_{\rm s,c}/c$ 
& sound velocity in corona  
& 0.91 or 0.65 \\
\enddata
\label{tab:parameter}
\end{deluxetable}

\begin{deluxetable}{ccccc}
\tablecolumns{5}
\tablewidth{0pc}
\tablecaption{Calculated models.}
\tablehead{
\colhead{Model}
& \colhead{plasma-$\beta$} 
& \colhead{$E_{\rm th,c}$}
& \colhead{$v_{\rm s,c}/c$}
& \colhead{$p_1$\tablenotemark{\dag}$/p_0$}
}
\startdata
 A & 10  & 1.0 & 0.91 & 0.54 \\
 B & 100 & 1.0 & 0.91 & 0.54 \\
 C & 10  & 0.1 & 0.65 & 0.054 \\
 D & 100 & 0.1 & 0.65 & 0.054 \\
\enddata
\label{tab:model}
\tablenotetext{\dag}{pressure of background corona at ($r_1$,0).}
\end{deluxetable}

We impose the absorbing inner boundary condition at a sphere $R_{1}=2
r_{\rm s}$.  The deviation from the initial values of $\rho$,
$p_{gas}$, $\mbox{\boldmath$v$}$, and $\mbox{\boldmath$B$}$ are damped
by a damping constant of:
\begin{equation}
{\cal A}\equiv {1-\tanh{\left[4(R-R_{1})/\Delta R\right]}\over 2}
\end{equation}
where $\Delta R=0.4 r_{\rm  s}$ is a transition width.  We calculate a
corrected quantity $q^{\rm new}$ at each time step from an uncorrected
quantity $q=q(r,\phi,z)$ by using the damping constant as:
\begin{equation}
q^{\rm new}=q-\left(q-q^{\rm init}\right){\cal A}
\end{equation}
where $q^{\rm init}$ is the initial value.

We impose a symmetric boundary condition on the equatorial plane.  On
the the symmetry axis (z-axis), $\rho$, $p_{\rm gas}$,
$\mbox{\boldmath$v$}_{z}$, and $\mbox{\boldmath$B$}_{z}$ are set to be
symmetric, while $\mbox{\boldmath$v$}_{r}$,
$\mbox{\boldmath$v$}_{\phi}$, $\mbox{\boldmath$B$}_{r}$, and
$\mbox{\boldmath$B$}_{\phi}$ are antisymmetric.  The outer boundary
conditions are free boundary condition where all the matter, magnetic
fields, and waves can transmit freely.  Accordingly, not only outflow
but also inflow from the outer boundary is permitted.  For more
details about the boundary conditions except the inner boundary, see
Matsumoto et al. (1996).

Hereafter, we normalize all the lengths, velocities, and density by
the Schwarzschild radius, $r_{\rm s}$, the speed of light, $c$, and
the initial torus density, $\rho_0 \equiv \rho(r_0,0)$, respectively.
Note that every term in each basic equation has the same dependence on
density if we express field strength in terms of plasma-$\beta$,
$B^{2}/8\pi = p_{gas}/\beta$.  Therefore, mass accretion rates can be
taken arbitrarily, as long as radiative cooling is indeed
negligible. This condition requires that accretion rates cannot exceed
some limit, over which radiative cooling is essential (Ichimaru 1977;
Narayan \& Yi 1995).

The basic equations are solved by the 3-D MHD code based on the
modified Lax-Wendroff scheme (Rubin \& Burstein 1967) with the
artificial viscosity (Richtmyer \& Morton 1967).  We use $200\times
32\times 200$ non-uniform mesh points.  The grid spacing is uniform
($\Delta r=\Delta z=0.16$) within the inner calculation box of $0\leq
r\leq 10$ and $0\leq z\leq 10$, and it increases by 1.5\% from one
mesh to the adjacent outer mesh outside this box up to $r\leq 20$ and
$z\leq 20$, and it increases by 3\% beyond that.  The entire
computational box size is $0\leq r\leq 100$ and $0\leq z\leq 100$.  We
simulate a full 360$^{\circ}$ domain (in comparison with HB02 and
INA03, who solved only a 90$^{\circ}$ wedge).

\section{MAGNETOHYDRODYNAMICAL ACCRETION FLOW AND JET}
\subsection{Emergence of a magnetic-tower jet}

We first display the overall evolution of Model A in Figure 1, in
which azimuthally averaged density contours with velocity vectors are
shown for 6 representative time steps.  Initially, dense,
geometrically thin disk is surrounded by a hot, tenuous background
corona (see the upper-left panel). When a disk material reaches the
innermost region close to the central black hole, upward motion of
gas is triggered (see the upper-middle panel),  which is driven by
increased magnetic pressure of accumulated toroidal fields (shown
later).  Matter is blown away upward by inflating toroidal fields (a
so-called magnetic tower) with a large speed, about several tenths of
$c$ (see the upper-right panel).  Some fraction of matter then goes
upward and leave the displayed region (note that this figure shows
only a part of our calculation box).  Eventually, the magnetic tower
stops vertical inflation and turns to shrink because of external
pressure by background corona (discussed later).  Accordingly, part of
blown up matter falls back towards the equatorial plane (see the
lower-left and lower-middle panels).  Finally, nearly quasi-steady
state realizes, when a geometrically thick structure persists (see the
lower-right panel).

\begin{figure*}[ht]
\centerline{\epsfxsize=\hsize \epsfbox{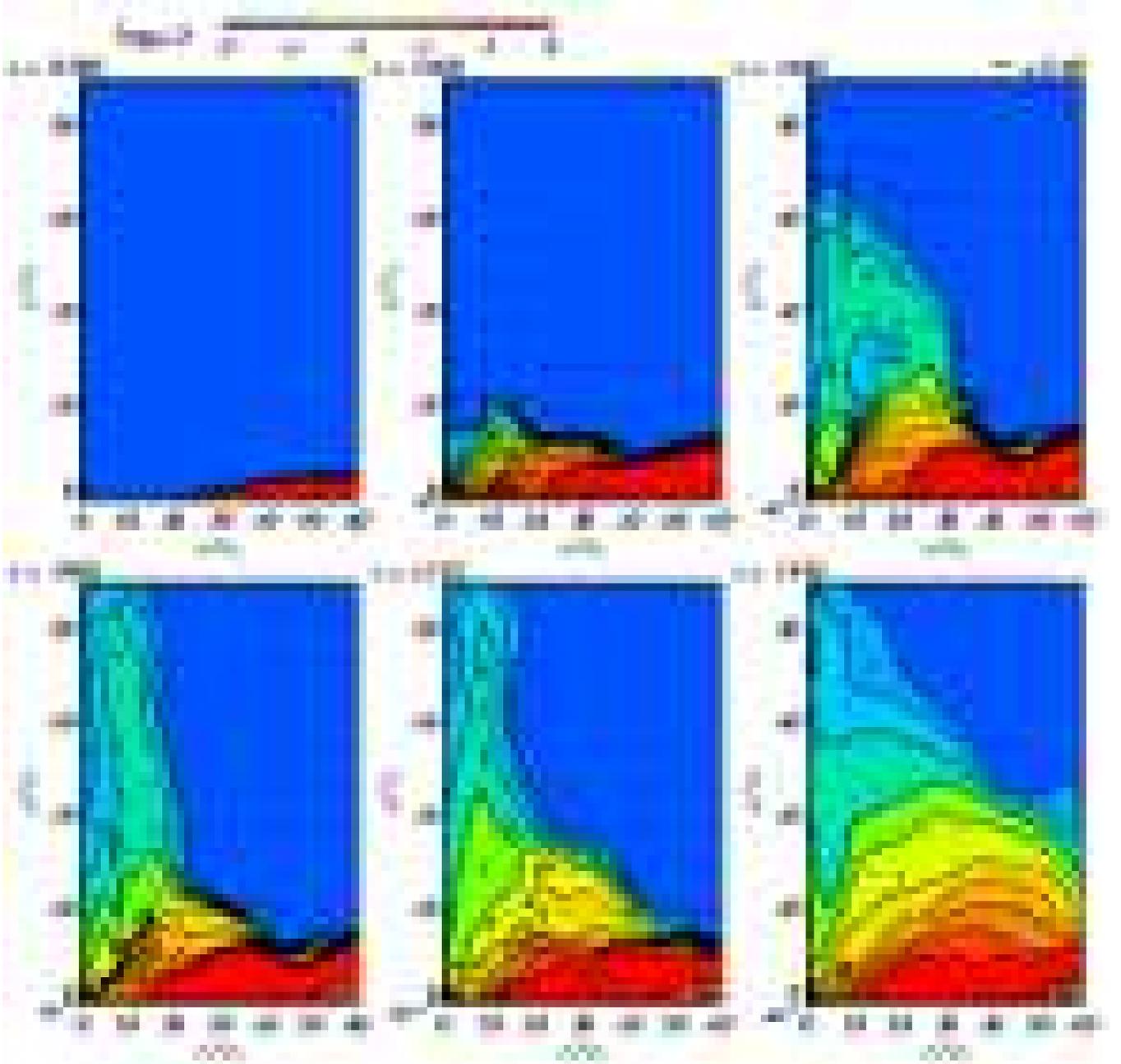}}
\caption{Time evolution of the density contours of Model A with
  velocity vectors being overlaid.  The density contours with colors
  are equally distributed on logarithmic scale with spacing being
  0.4.  The elapsed times are indicated in the upper-left corner of
  each panel.  The unit of time is $r_{\rm s}/c = 10^{-4}
  (M/10M_\odot)$ s.  A magnetic jet first grows at $t = 1000 \sim
  1800$ but loses power at later times, producing more spherical
  density distribution. }
\label{fig1:eps}
\end{figure*}

We can roughly divide the entire evolution of a rotating torus
initially threaded by poloidal fields into two distinct phases: a
transient, extended jet phase (hereafter, Phase I) and a subsequent
quasi-steady phase (or Phase II).

Figure 2 displays more detailed structure in Phase I of Model A.  The
left panel shows the contours of the plasma-$\beta$ on the logarithmic
scale.  We can easily distinguish jet regions which have blue colors
and thus are in low-$\beta$, and ambient red regions which are in
high-$\beta$.  Most of matter inside the jet is blown outward,
although downward motion is also observed partly, especially near the
rotation axis.  Total pressure (a sum of gas pressure and magnetic
pressure) decreases with an increase of height in nearly
plane-parallel fashion, as is shown by the contours by black solid
lines.  It is interesting to note that total pressure inside the jet
is nearly uniform in the horizontal direction at $z > 20$, indicating
that the magnetic tower is balanced with the ambient gas pressure.
Without external pressure by background corona, a magnetic tower would
expand along the directions of 60$^\circ$ tilted from the rotation
axis (Lynden-Bell \& Boily 1994).  Model A shows that a magnetic tower
evolves nearly in the vertical direction, which is made possible
because of substantial external pressure (Lynden-Bell 1996).

\begin{figure*}[ht]
\centerline{\epsfxsize=\hsize \epsfbox{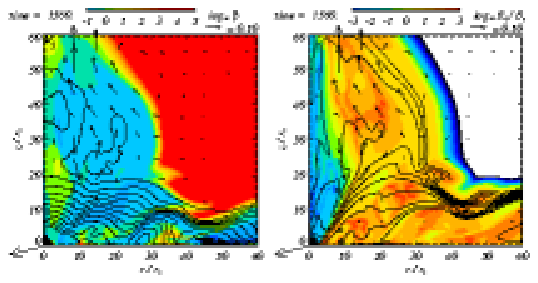}}
\caption{Color contours of plasma-$\beta$ (left) and magnetic pitch
  (right) in Phase I of Model A, respectively, overlaid with velocity
  vectors and total pressure (left) and density (right) contours with
  black solid lines in logarithmic scale.  Contour level of total
  pressure is $P_{tot}(N)=-6.0+0.15 N$ where $N$ is integer from 0 to
  $14$ while that of density is same as fig 1.}
\label{fig2:eps}
\end{figure*}

The right panel of figure 2 shows the contours of the
$B_{\rm\phi}/B_{\rm p}$.  This figure is useful to check which of
toroidal or poloidal fields are dominant in which region.  We can
immediately understand that the core of the jet has a blue color,
implying poloidal components being more important, while the
surrounding zone of the jet has a orange color, indicating dominant
toroidal field components.  Obviously, toroidal fields can be easily
generated, as matter drifts inward, owing to faster rotation and
larger shear at smaller radii.

Such different field configurations near the surface and in the core
of the jet is more artistically visualized in the three-dimensional
fashion in figure 3.  Here, thick red lines indicate magnetic field
lines and thick green-to-blue lines represent streamlines with
indication of velocities by the color; the blue, green, and yellow
represent velocities of $\sim 0.1$, 0.3, and 0.4 (in the unit of $c$),
respectively.  We can understand why magnetic field configuration of a
sort that we encounter here is called as ``a magnetic tower.''  Volume
inside the tower is mostly occupied by tightly wound toroidal fields.
At the same time, we also observe vertical field lines in the core of
the jet.  As time goes on, this magnetic tower inflates vertically due
to the growing magnetic pressure by the accumulated toroidal fields.
Accordingly, the jet is also driven by enhanced magnetic pressure of
the magnetic tower.  Matter attains velocity up to $\sim 0.1 - 0.2 c$
at the top.

\begin{figure*}[ht]
\centerline{\epsfxsize=\hsize \epsfbox{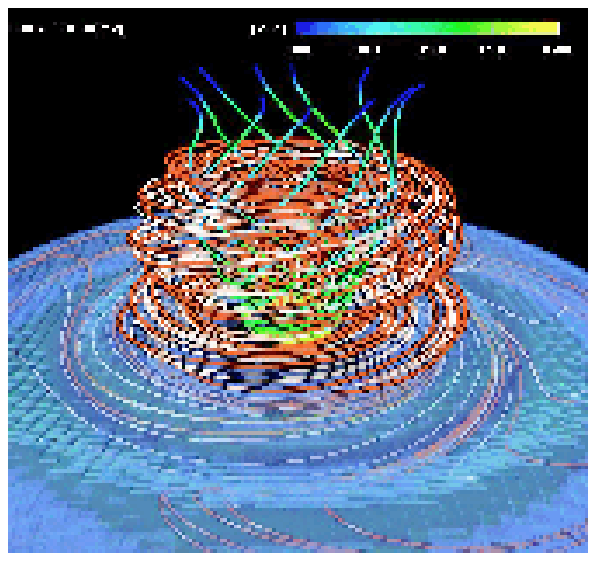}}
\caption{Perspective view of magnetic fields lines in Phase I of Model
  A.  Thick red (or thin white) lines indicate magnetic field lines
  which are anchored to the innermost (somewhat outer) zones at
  $(r,z)=(1,1.5)$ [$(r,z)=(56,10)$], respectively.  Thick green lines
  denote the streamlines of velocity vectors integrated from
  $(r,z)=(8.5,7)$, whereas the color bar indicate the velocity.
  Light-blue shaded region indicate the isovolume of the density
  ($\rho=0.025\rho_{0}$).  Accumulated toroidal fields emerging from
  the disk produce a magnetic tower, thereby driving an MHD jet.  Jet
  material is surrounded by toroidal magnetic fields, whereas poloidal
  (vertical) fields dominate inside the jet.}
\label{fig3:eps}
\end{figure*}

We have checked how much fraction of energy leaves the calculation box
in which form (see Table 3).  The values in the table are shown by the
percentage of the sum of time-averaged fluxes listed in the column at
each phase.  Generally, the energy carried by jets does not dominate
the total energy loss from the system.  In Model A, for example, the
outflow energy is only $\sim 0.5\%$ of the total energy loss.

\begin{deluxetable}{ccrrrrr}
\tablecolumns{7}
\tablewidth{0pc}
\tablecaption{Fraction of flux.}
\tablehead{
\colhead{flow}
& \colhead{flux\tablenotemark{\dag}}
& \colhead{A/I [\%]}
& \colhead{A/II [\%]}
& \colhead{B [\%]}
& \colhead{C [\%]}
& \colhead{D [\%]}
}
\startdata
outflow & {$c\langle\mbox{\boldmath$E$}\times\mbox{\boldmath$B$}\rangle_{z}\tablenotemark{\ddag}~/4\pi$}
& {$0.13$} &{$0.00$}
& {$0.00$}
& {$3.14$}
& {$0.53$} \\
        & {$\langle (\rho v^{2}/2)\mbox{\boldmath$v$}_z \rangle$ }
& {$0.14$} & {$0.00$}
& {$0.00$}
& {$8.99$}
& {$0.69$} \\
        & {$\langle \gamma p/(\gamma -1)\mbox{\boldmath$v$}_z\rangle$ }
& {$0.27$} & {$0.00$}
& {$0.00$}
& {$1.21$} 
& {$0.21$} \\
inflow & {$c\langle\mbox{\boldmath$E$}\times\mbox{\boldmath$B$}\rangle_{in}\tablenotemark{*}~/4\pi$}
& {$0.03$} & {$0.03$} 
& {$0.05$}
& {$1.0$} 
& {$0.22$} \\
        & {$\langle (\rho v^{2}/2)
  \mbox{\boldmath$v$}_{in}\rangle$} 
& {$29.11$} & {$29.25$} 
& {$28.34$}
& {$28.29$}
& {$28.19$} \\
         & {$\langle \gamma p /(\gamma -1)
  \mbox{\boldmath$v$}_{in}\rangle$}
& {$30.99$} & {$31.13$}
& {$30.15$}
& {$34.94$}
& {$30.20$} \\
        & {$\langle \rho\psi\mbox{\boldmath$v$}_{in}\rangle$}
& {$39.42$} & {$39.60$}
& {$41.45$}
& {$24.96$} 
& {$40.40$} \\
\enddata
\label{tab:e-flux}
\tablenotetext{\dag}{The energy and enthalpy flux passing though
  $z=50$ is azimuthally averaged and is integrated over $5\leq r\leq
  30$, while that of the inflow is integrated inside the sphere
  $R=2.2$.}
\tablenotetext{\ddag}{The subscript $z$ indicates the vertical
  (z-) component of a vector.}
\tablenotetext{*}{The subscript $in$ indicates the
  $\mbox{\boldmath$e_{in}$}$-component of a
  vector where $\mbox{\boldmath$e$}_{in}= - (\mbox{\boldmath$e$}_{r}
  \cos\theta + \mbox{\boldmath$e$}_{z} \sin\theta)$ is a unit vector,
  $\sin\theta / \cos\theta = z/r$, and $\mbox{\boldmath$e$}_{r}$
  is the radial (r-) component of a unit vector.}
\end{deluxetable}

\subsection{Quasi-steady MHD flow}

Eventually the initial jet loses its power, and the system enters a
quiet, quasi-steady state.  First, we show in figure 4 the radial
distributions of representative physical quantities in both of Phase I
(by the thick solid lines) and Phase II (by the thick dashed lines).
The adopted quantities are density, radial velocity, pressure,
specific angular momentum, sound velocity, and mass accretion rate,
$\dot{M}\equiv -\int_{0}^{H}2\pi r\rho v_r dz$ with $H =1.0$ being
fixed.  These quantities are azimuthally and vertically averaged over
$0\leq \phi\leq 2\pi$, $0 \leq z \leq 1$ except the mass accretion
rate which is a vertical integral.

\begin{figure}[h]
\centerline{\epsfxsize=\hsize \epsfbox{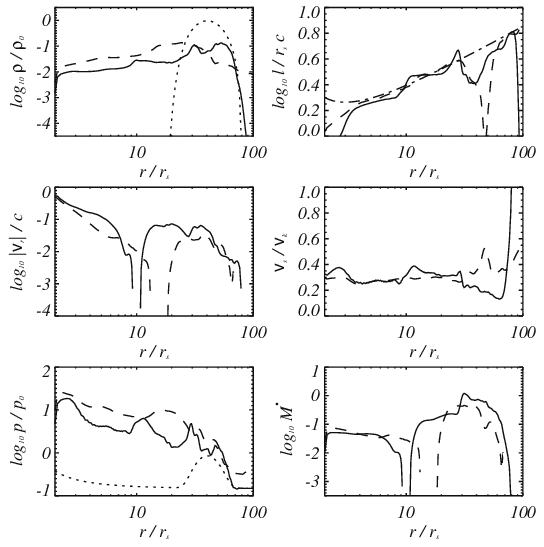}}
\caption{Radial structure of the flow of Model A on the equatorial
  plane in Phase I ($t=1560 \; r_{\rm s}/c$; transient extended jet
  phase) by the solid curves and in phase II ($t=3910 \; r_{\rm s}/c$;
  subsequent quasi-steady phase) by the dashed curves,
  respectively. From the top left to the right bottom, density, radial
  velocity, pressure, specific angular momentum, sound velocity, and
  mass accretion rate are plotted.  The dotted lines indicate initial
  conditions for density and pressure, while a dash-dotted line
  indicates a Keplerian specific angular momentum.}
\label{fig4:eps}
\end{figure}

Roughly speaking, these quantities show power-law relations; i.e.,
$\rho \propto r$ at small radii, while $\rho \propto r^{-1}$ at large
radii, $v_r \propto r^{-1/2}$ except at the inner outflowing part,
$\ell$ (specific angular momentum) $\propto r^{0.5}$, meaning a nearly
Keplerian rotation,  $c_{\rm s}$ (sound velocity) $\propto r^{-1/2}$,
and $\dot M \sim $const. in space in the inner region.  All of these
relations resemble closely those of the previous work (i.e., Stone \&
Pringle 2001, HB02).  Note that our results are consistent with a hot,
thick, near-Keplerian disk, and a subthermal magnetic field.  The
density distribution is inconsistent with ADAF models, since ADAF
shows a steeper density profile, approximately $\rho \propto
r^{-3/2}$.

The most interesting feature in the present simulations is found
in figure 5, which illustrates contours of the total pressure (by
black solid lines) and plasma-$\beta$ (by color contours) in the left
panel and those of density (by black solid lines) and
$B_{\phi}/B_{p}$ (by color contours) in the right panel in Phase II of
Model A calculations.  We find that total pressure distribution is not
plane-parallel and the plasma-$\beta$ distribution has inhomogeneous
structure around the central black hole.

\begin{figure*}[ht]
\centerline{\epsfxsize=\hsize \epsfbox{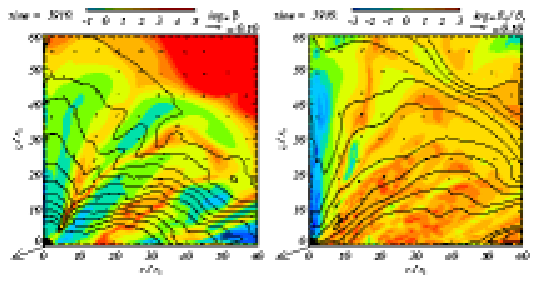}}
\caption{Same as Fig. 2 but in Phase II of Model A.}
\label{fig5:eps}
\end{figure*}

Interestingly, high-$\beta$ ($\sim 100$) plasmas (indicated by orange
color in the left panel) distributed in the radially extended zone
which is tilted by $\sim 30^\circ$ from the $z$-axis (in the
perpendicular direction to the iso-pressure contours) are sandwiched
by moderately low-$\beta$ ($< 1$) plasmas indicated by blue color.
There is another high-$\beta$ region extending from point
($r,z$)$\sim$($15,0$) to the upper left region.  We also notice that
poloidal (or toroidal) fields are dominant in these high-$\beta$
(low-$\beta$) regions, as is indicated by orange (green or blue)
colors in the right panel.  As a consequence, mass accretion onto the
central black hole takes place mainly through the equatorial plane at
$r < 20$.  We also notice that poloidal (vertical) fields are still
evident along the rotation axis, along which downward motion of gas is
observed.

We find that equi-$\beta$ contours slightly change with time.  The
  accretion flow is inherently time dependent, as it is driven by MHD
  turbulence.  There are two possibilities for the origin of time
  dependence in this case.  One is the axisymmetric mode of MRI
  (so-called channeling mode).  Since the structure of quasi-steady
  flow is nearly-axisymmetric, the channeling mode of MRI can grow in
  our simulations.  As a result, high-$\beta$ plasma is created
  between the low-$\beta$ plasmas with opposite directions of the
  flow.  Another possibility is magnetic buoyancy.  If we assume
  pressure balance among neighboring high- and low-$\beta$ regions,
  matter density should be high in the former, since magnetic pressure
  is less in the former.  This gives rise to magnetic buoyancy for the
  latter, inducing its rising motion in the direction opposite to that
  of gravity.  Such a rising magnetic loop was simulated by Turner,
  Bodenheimer, and R{\' o}{\.z}yczka (1999), who found that the rising
  loop evolves into a jet (see also KMS02 for global simulations).  In
  both cases, magnetic reconnection can be triggered.  We find that
  the directions of toroidal fields are anti-parallel across
  high-$\beta$ region.  Thus, magnetic reconnection can take place in
  this high-$\beta$ region and high-velocity plasmoid can be ejected
  from this site.   If this is the case, high-$\beta$ region may move
  outward.  Velocity vectors plotted in this figure show both of
  outward and inward motions, although their absolute values are
  small.  The situation is rather complex.  To make clear what is
  happening in the quasi-steady state, we need more careful studies.

It will be highly curious to see how magnetic field configuration
changes after the jet eruptions. In figure 6 we plot the perspective
view of the magnetic fields lines and matter trajectories in Phase II
of Model A.  Field line configurations are distinct from those in
Phase I.  Field lines which are connected to the innermost part of the
flow (indicated by red thick lines) now show much simpler structure
than in Phase I and are dominated by poloidal component, implying that
toroidal fields found in Phase I have already been blown away.  Note,
however, that the surrounding region is still dominated by toroidal
fields, as are indicated by white thick lines.  As a consequence, the
magnetic tower is not clearly seen.

\begin{figure*}[h]
\centerline{\epsfxsize=\hsize \epsfbox{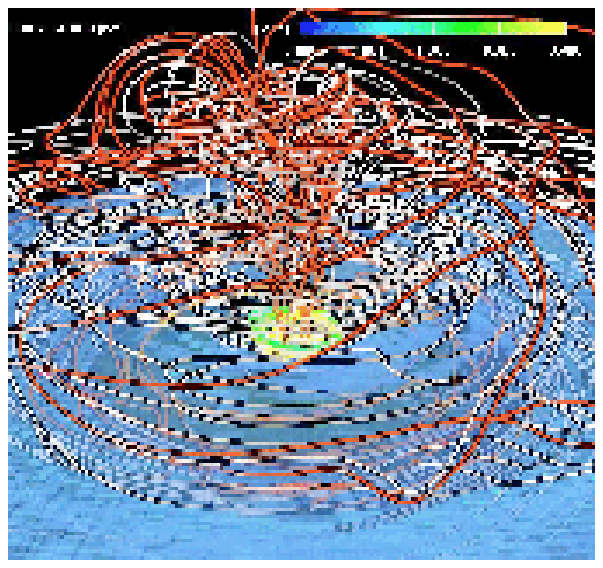}}
\caption{Same as Fig. 3 but in Phase II of Model A.  Magnetic fields
  lines represented by thick red lines are more confined, compared
  with those in figure 3.  Thin white lines (representing the fields
  lines emerging from somewhat outer zones) look similar to those in
  figure 3.  Thick green lines indicating the streamlines of velocity
  vectors only appear near the black hole and are not evident.}
\label{fig6:eps}
\end{figure*}

In order to see more clearly how toroidal and poloidal fields are
generated and accumulated with time in the disk and the corona,
we plot the strength of each component as a function of radius at
several time steps in figure 7.  Toroidal fields increase in strength
by the shear and dominates in the entire region (at $t=1200$), but they
are blown away from the disk by the emergence of a jet and thus
decrease in strengths (at $t=1560$), whereas poloidal fields are
generated due to the emergence of the magnetic tower from the
disk. After the substantial ejection of matter and toroidal fields,
the axial part of the corona $r < 7$ is dominated by poloidal fields.
Toroidal fields can be blown off with jet material, but poloidal
fields created by vertical inflation of toroidal fields should
remain.  Note that, since no continuous injection of mass and fields
are assumed in our calculation, the increase of the poloidal fields in
the inner zone is eventually saturated.  If we would add more fields,
they would continuously grow.

\begin{figure}[h]
\centerline{\epsfxsize=0pt \epsfbox{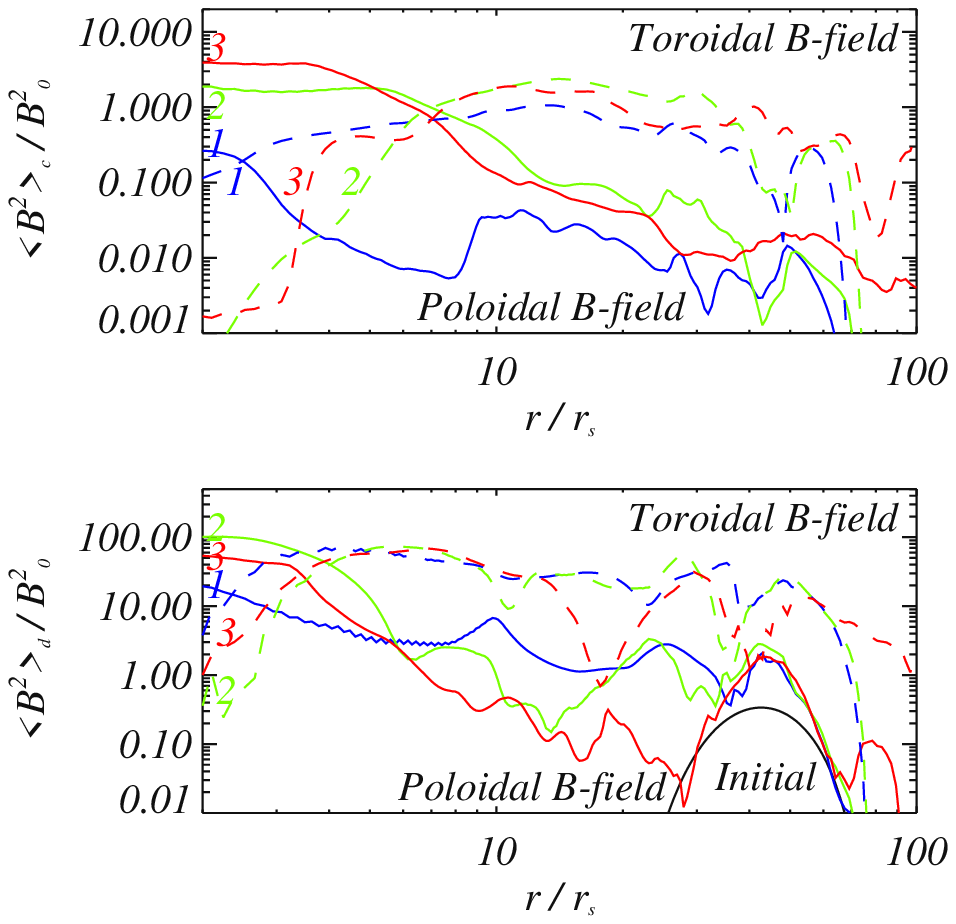}}
\caption{Changes of the magnetic energy density in the disk
    (bottom panel) and in the corona (upper panel) from their initial
    values as functions of radius for Model A.  Each panel presents
    the magnetic energy density of poloidal (solid lines) and that of
    toroidal fields (dashed lines).  The magnetic energy in the disk
    (or in the corona) is integrated from $z=0$ to $z=5$ (from $z=5$
    to $z=100$).   The elapsed times are $1200$, $1560$, and $3910$
    (in $r_{s}/c$), corresponding to the labels 1, 2, and 3,
    respectively.}
\label{fig7:eps}
\end{figure}

We finally plot the energetics of Model A in figure 8.  Kinetic energy
always dominate others in both phases.  That is, magnetic fields are
not strong enough to suppress accreting motion of the gas.  There is
no evidence of huge buildup of net field which would inhibit accretion
onto the central black hole in our simulations, unlike the picture
obtained by INA03.

\begin{figure}[h]
\centerline{\epsfxsize=0pt \epsfbox{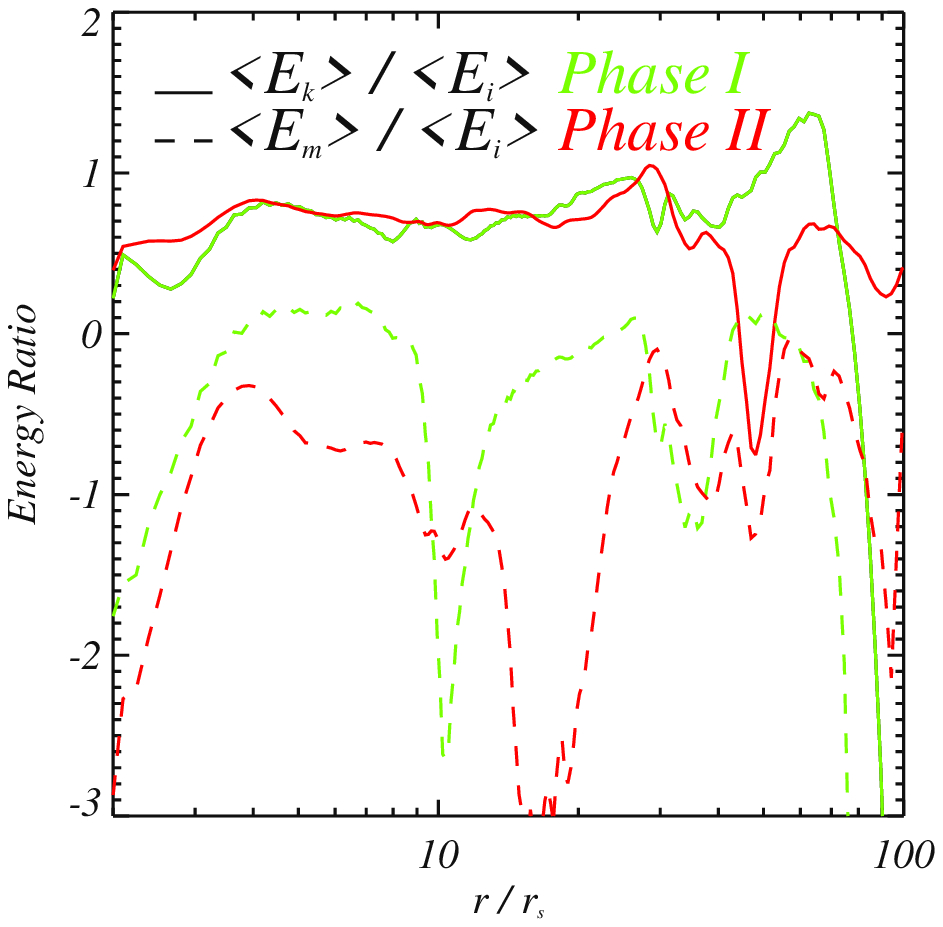}}
\caption{Energetics of Model A.  The ratio of kinetic energy ($E_{\rm
    k}$) to thermal energy ($E_{\rm i}$) and the ratio of magnetic
  energy ($E_{\rm mag}$) to thermal energy ($E_{\rm i}$) are plotted
  as functions of radius in Phases I ($t=1560 \; r_{\rm s}/c$) and II
  ($t=3910 \; r_{\rm s}/c$), respectively.  Kinetic energy always
    dominate over others by about one order of magnitude.}
\label{fig8:eps}
\end{figure}

\subsection{Cases of weak initial fields and/or low external pressure}

Before closing this section, let us see some parameter dependence of
our results.  In Models B -- D, we vary the initial magnetic-fields
strengths and/or external pressure by background corona.  To help
understanding different conditions, we plot the initial radial
distribution of pressure (gas and magnetic) in figure 9 for each
model.  We see that external pressure is comparable to initial torus
pressure in Models A and B, while the external pressure is totally
negligible in Models C and D.

\begin{figure}[h]
\centerline{\epsfxsize=0pt \epsfbox{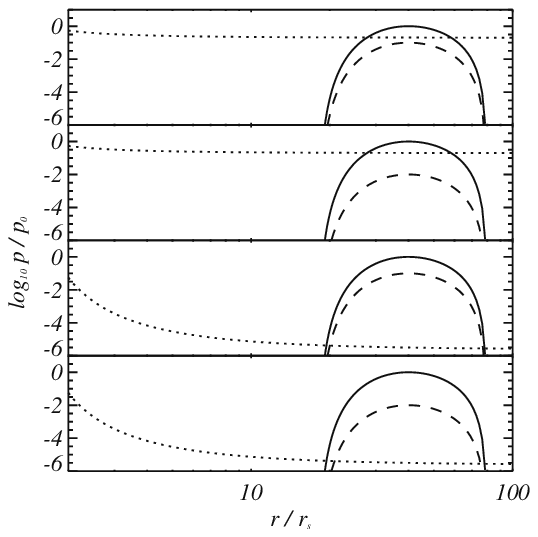}}
\caption{Initial pressure distributions for Model A, B, C, and D
 from the top.  The solid and dashed lines represent the gas and
 magnetic pressures at $z=0$, respectively, where as the dotted lines
 represent external pressures by background corona extrapolated to the
 $r$-axis ($z=0$; see equation \ref{Pcorona}).  All values are
 normalized by the pressure at the density maximum of the torus ($r$,
 $z$)=($r_{0}$,0)}
\label{fig9:eps}
\end{figure}

The later evolution is shown in figure 10.  First, the case of
initially weaker magnetic fields (Model B) is displayed in the second
row of figure 10.  Importantly, we do not see a strong outflow.  The
absence of an obvious jet can be understood in terms of pressure
balance.  Since magnetic field strength is by one order of magnitude
smaller in Model B than that in Model A, the initial magnetic pressure
is much smaller than external pressure (see figure 9).  Magnetic
pressure can grow as disk dynamo works, but magnetic pressure is not
yet large enough to overcome external pressure at the moment that the
disk material reaches the innermost region.  Note that the time needed
for the torus material reaches the center is much longer in Model B
than in Model A, because it takes a longer time for fields to be
strong enough to promote rapid accretion.  Since the magnetic pressure
inside the disk is comparable to the background pressure in Model A
initially, it is easy to overcome the background pressure.  Hence,
eruption of a magnetic jet is possible.  Nevertheless, it is
interesting to note that the subsequent states look similar among
Models A and B.  This indicates that Phase II structure is not
sensitive to the strengths of initial magnetic fields, although it is
not yet clear that the results may or may not depend on the initial
magnetic field orientations.

\begin{figure*}[h]
\centerline{\epsfxsize=\hsize \epsfbox{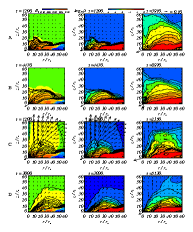}}
\caption{Summary of the results of 4 calculated models.
From the left to the right, contours of toroidal field strengths
and of density in different time are displayed for Models A, B, C, and
D from the top.  Arrows show the velocity vectors.  Initial magnetic
fields strengths are weaker in Models B and D (in which $\beta = 100$
while $\beta = 10$ otherwise), while external pressure is 10 time
weaker in Models C and D.}
\label{fig10:eps}
\end{figure*}

The cases of weaker external pressure (Models C and D) are illustrated
in the third and fourth rows of figure 10.  We find stronger, wider,
and longer-standing (nearly persistent) jet in these models.  This
demonstrates that duration of a jet, as well as its width, are
sensitive to how large external pressure is.  Again, the flow
structure in Phase II looks similar among Models C and D, but unlike
cases with stronger external pressure (Models A and B) there is a
conical low-density region present near the symmetric ($z$-) axis.
This is due to the ejection of mass from the innermost part via
jets.

We find in table 3 that the total energy output by the jet amount to
$\sim 10$\% of the total energy loss at maximum in Model C (with
strong initial magnetic field and low external pressure).  Except this
model, total energy output is negligible.  How much fraction of energy
can be carried away from the black hole system thus depends critically
on external pressure and field strength, and it is not always possible
to achieve a high efficiency of energy extraction by jets.

Finally, we summarize our simulation results in the schematic picture
  (see figure 11).  In figure 11, high-$\beta$ plasma with highly
  turbulent motion locates at the equatorial plane.  Magnetic field
  generated by the dynamo action taking place inside the disk emerges
  upward and create the disk corona with intermediate-$\beta$ plasma.
  In the inner region of the disk, the emerging magnetic field erupts
  into the external corona and deform into the shape of the magnetic
  tower.  Such magnetic tower ejects disk materials upward by its
  magnetic pressure and creates the magnetic tower jet.  The magnetic
  tower jet collimates at the radius where the magnetic pressure of
  the tower is balanced with the external pressure.

\begin{figure*}[h]
\centerline{\epsfxsize=\hsize \epsfbox{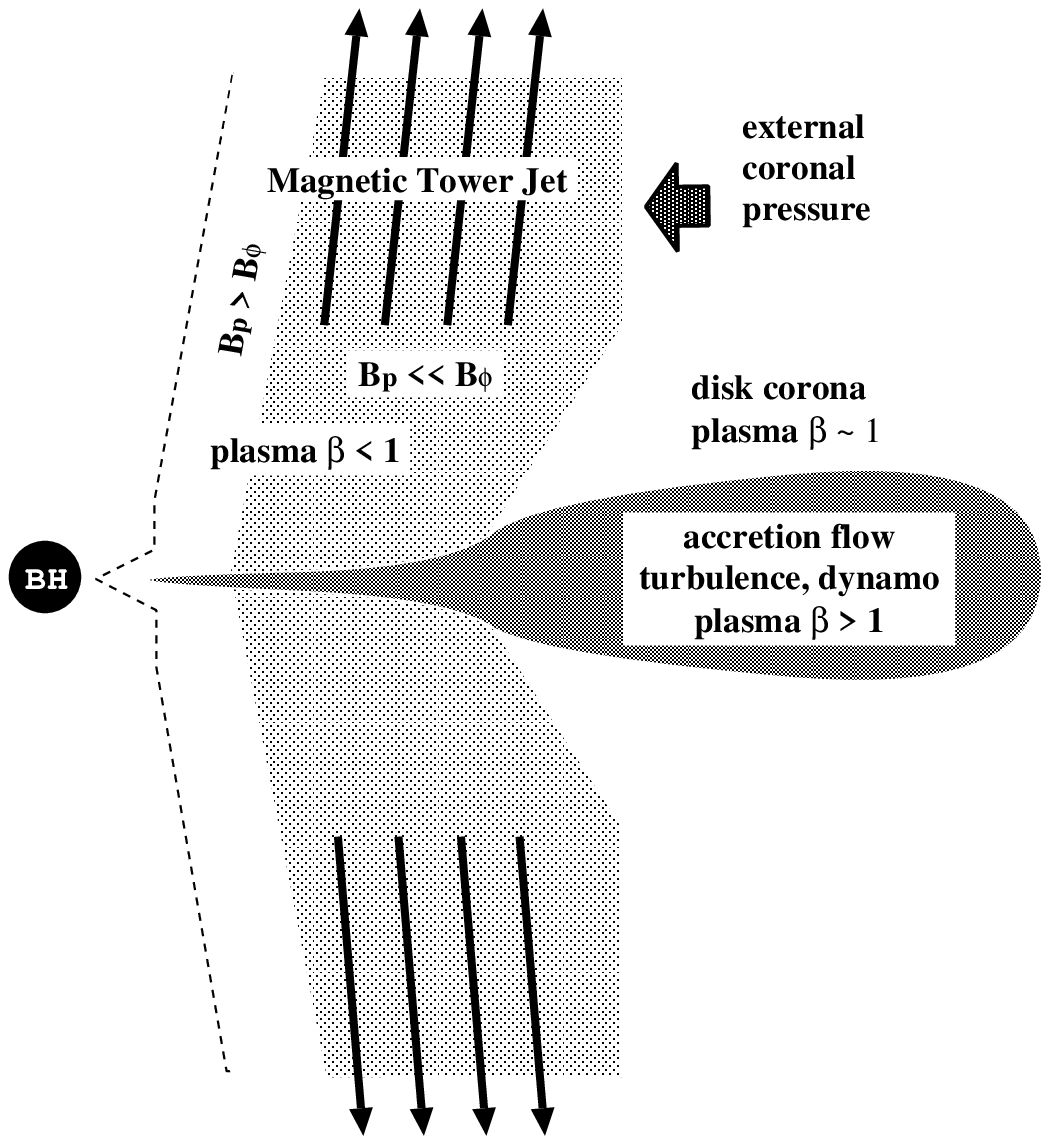}}
\caption{Schematics of the magnetic tower jets.}
\label{fig11:eps}
\end{figure*}

\section{DISCUSSION}
\subsection{Brief Summary}

We performed 3-D MHD simulations of the radiatively inefficient
rotating accretion flow around black holes.  We examined the case with
initially poloidal field configurations with $\beta =10$ and 100.
When the bulk of torus material reaches the innermost region close to
the central black hole, a magnetic jet is driven by magnetic pressure
asserted by accumulated toroidal fields (magnetic tower).  
The fields are mostly toroidal in the surface regions of the jets, 
whereas poloidal (vertical) fields dominate in the inner core of the
jet.  The collimation width of the magnetic jet depends on external
pressure; the more enhanced the external pressure is, the more
collimated the jet is.  Non-negligible external pressure tends to
suppress the emergence of the MHD jets.

The jet outflow is, however, a transient phenomenon, unless the
external pressure is negligible.  After several dynamical timescales
at the radius of the initial torus, the jet region shrinks and the
flow structure completely changes.  After the change, the flow is
quasi-steady and possesses complex field configurations.  

The nature of the quasi-steady state differs from that of INA03.  We
have seen that poloidal fields are dominant near the $z$-axis and that
the black hole is mostly surrounded by marginally low-$\beta$ plasma
with toroidal fields being dominant.  This view is different from that
simulated by INA03, who found a magnetosphere composed of a
significantly low-$\beta$ ($\sim 10^{-3}$) plasmas with poloidal
fields and that mass is not allowed to directly pass into the
magnetosphere.  Instead, mass accretion occurs along the radial narrow
streams with slow rotation.  We will discuss in \S 4.3 what the origin
of difference would be.

\subsection{Magnetic Tower Jet}
The acceleration mechanism of the MHD jet, starting with a disk
threaded with large-scale, external vertical fields, has been studied
extensively by many groups.  It has been discussed that in the
magneto-centrifugally driven jet, the poloidal magnetic field is much
stronger than the toroidal magnetic field in the surface layer of the
disk or in the disk corona, where plasma-$\beta$ is low (e.g.,
Blandford and Payne 1982; Pudritz and Norman 1986; Lovelace et
al. 1987).  In this case, magnetized plasma corotates with the disk
until the Alfv\'{e}n point, beyond which toroidal field starts to be
dominated and hence collimation begins by its magnetic pinch effect.
Blandford-Payne (1982) mechanism can be applied to this kind of jet
for determining the launching point of the jet.

On the other hand, there is another kind of magnetically driven jet,
in which the toroidal magnetic field is dominated everywhere (e.g.,
Shibata and Uchida 1985; Shibata, Tajima, Matsumoto 1990; Fukue 1990;
Fukue, Shibata, Okada 1991; Contopoulos 1995). In this case,
Alfv\'{e}n point is embedded in the disk or there is no Alfv\'{e}n
point, and the collimation due to toroidal field begins from the
starting point of the jet. Blandford-Payne mechanism cannot be applied
to such jet.

Situation is similar, if magnetic loops are anchored to the surface of
the differentially rotating disk or that of the central object.  MHD
jet generated from the weak, localized poloidal field in the disk
(Turner et al. 1999, KMS02) belong to the toroidal field dominated
jet.  Moreover, MHD jet generated from the dipole magnetic field is
simulated by Kato, Hayashi, \& Matsumoto (2004).  They calculated time
evolution of a magnetic tower jet, which is an extension of
Lynden-Bell (1996)'s magnetic tower, and is also the toroidal field
dominated jet.

Our magnetic tower jet is also consistent to the toroidal field
dominated jet, since the magnetic tower is made of the toroidal field
generated by dynamo action within the disk.  It has often been argued
that such a toroidal field dominated jet is very unstable for kink
instability in real 3D space, and cannot survive in actual situation
(e.g., Spruit, Foglizzo, Stehle 1997).  However, our 3D simulations
showed, for the first time, that such toroidal field dominated jet
survive at least for a few orbital periods of the disk and hence can
be one of promising models for astrophysical jets.

By means of the magnetic tower jets, one may ask what the condition
  for the formation of the toroidal field dominated jet is.  The
  formation of magnetic jets demand that the maximum magnetic pressure
  $p_{\rm mag}^{\rm max}$ generated within the disk must exceed the
  external pressure $p_{\rm ext}$.  Since the growth of toroidal
  fields due to MRI is saturated under a certain value (e.g.,
  $\beta_{min}\sim 100$ within the disk; see Matsumoto et al. 1999;
  Hawley 2000; INA03), we have a condition for the formation of the
  jet as follows:
\begin{equation}
p_{\rm mag}^{max}=p_{\rm disk}/\beta_{min}>p_{\rm ext},
\end{equation}
where $p_{\rm disk}$ is the pressure of the disk.  Our conclusion is
equivalent to that of KMS02 for the case of isothermal background
corona, in which they argued that if the coronal density is very high,
it prevents the ejection of magnetic loop.

If external pressure is very small, compared with initial torus
(magnetic) pressure, magnetic fields can more easily escape from the
disk than otherwise.  The result is a strong, long-persistent jet, as
we saw in Models C and D.  Although the collimation is not clear in
the plot with limited box, collimation does occur but at large radii
($> 60 r_{s}$) in these models, since, as the outflow expands magnetic
pressure decreases and eventually becomes comparable to the external
pressure.  We thus call it a jet, not a big wind, and the width of the
jet depends on the external pressure.

If external pressure is very large and by far exceeds
the maximum magnetic pressure inside the disk, conversely, the
emergence of magnetic fields out of a disk is totally inhibited, the
situation corresponding to Model B.  There will be no jets.  Only when
external pressure is mildly strong, comparable to the maximum magnetic
pressure, we can have a transient eruption of a collimated magnetic
jet (see Model A).

It is important to note, in this respect, the works by Lynden-Bell.
Lynden-Bell \& Boily (1994) studied self-similar solutions of
force-free helical field configurations, finding that such
configurations expand along a direction of 60 degrees away from the
axis of helical field.  Modified model by Lynden-Bell (1996) asserts
that a magnetic tower can be confined by external pressure.  Here, by
3-D MHD simulation, we successfully show that the formation process of
magnetic tower, which is stable for a few dynamical timescale at the
density maximum of the initial torus.  Our 3-D magnetic tower solution
is basically the same as that he proposed.

Unfortunately, there were not always detailed descriptions
regarding external pressure given in previous MHD simulation papers,
although the presence of external pressure is indispensable to perform
long-term simulations.  INA03 claimed that even with continuous supply
of mass and fields, bipolar jet (or outflow) phase is transient.
Their case seems to correspond to the case with relatively strong
external pressure (i.e., our Model A).  If this is the case in actual
situations, we expect strong jets only when a burst of large accretion
flow takes place.  This may account for the observations of a
micro-quasar, GRS1915+105, which recorded big (superluminal) jets
always after sudden brightening of the system (Klein-Wolt et
al. 2002).

It is interesting to discuss the collimation mechanisms of the jet.
We checked the force balance, confirming that it is external pressure
that is balanced with magnetic pressure at the boundary between the
jet and the ambient corona.  Together with our finding that jet widths
depend also on the value of external pressure, we are led to
conclusion that for jet collimation, external pressure play important
roles in the present case.

Furthermore, the total pressure of magnetically dominated ``corona'',
created above the disk due to the eruption of the magnetic tower jet
in early phase, could be greater than that of the background pressure
(hereafter we call it as ``disk corona'' in order to distinguish from
background corona; see also two-layered structure of accretion flow as
presented by HB02).  Thus we expect that the collimation widths of a
successive jet would be smaller than that in the case of Model A.

To sum up, we demonstrate the important role of external pressure
asserted by ambient coronal regions on the emergence, evolution, and
structure of a magnetically driven jet.  We need to remark how
feasible the existence of a non-negligible external pressure is in
realistic situations.  Fortunately, we have a good reason to believe
that ambient space of accretion flow system may not be empty.  If
there is magnetic dynamo activity in the flow, magnetic fields will be
amplified and eventually escape from the flow forming a magnetically
confined corona (Galeev, Rosner, \& Vaiana 1979).  Then, mass
can be supplied from the flow to a corona via conduction heating of
the flow material.  In fact, the density of the disk corona is
basically explained by the energy balance between conduction heating
and evaporation cooling of disk chromosphere (Liu, Mineshige, \&
Shibata 2002).  Thus, it is reasonable to assume that ambient space
surrounding a magnetized flow is full with hot tenuous plasmas as a
result of previous magnetic activity of the disk and that the hot
corona exerting external pressure on the flow with comparable
magnitude.

Finally, we remark on the comparison with the simulations by
HB02.  They also calculated the evolution of a torus threaded with
poloidal fields, adopting similar initial conditions as we did.  But
there is a distinction; we find a significantly low-$\beta$ jet,
whereas HB02 obtained a high-$\beta$ jet (see their figure 1).  

In contrast to HB02, our jet is driven by the magnetic pressure inside
magnetic tower.  Although we found the outflow along the funnel
barrier inside the core of magnetic tower in Model C and D (fig 10),
the outflow is not driven by the gas pressure but the magnetic
pressure.  Furthermore, we note that averaged plasma-$\beta$ along the
funnel barrier where the poloidal field dominates the toroidal field
is of the order of unity (fig 5).  In comparison with HB02, the
discrepancy can be explained by the difference of external
pressure. For instance, in Model C and D, there are persistent jets
which look like uncollimated winds reported in HB02.  This is because
that the external pressure was too low to confine the magnetic tower
inside the computational box.

\subsection{Steady-State Picture of MHD Accretion Flow}

The flow structures in the quasi-steady phase which we obtained in our
simulations look similar to that of previous work except INA03.  There
is a big distinction; INA03 claimed the formation of a significantly
low-$\beta$ (with $\beta \sim 10^{-3}$) magnetosphere dominated by a
dipolar magnetic field.  Since any motion passing through the
magnetosphere is inhibited, accretion occurs mainly via narrow
slowly-rotating radial streams.  We, in contrast, find the formation
of a $\beta \sim 1$ region (on average) which is dominated by toroidal
magnetic fields, although the structure is quite inhomogeneous and is
a mixture of high- and low-$\beta$ zones. Importantly, the fields are
not too strong to inhibit gas accretion, since magnetic field strength
is amplified by the dynamo action inside the disk due to MRI and the
spatially averaged plasma-$\beta$ inside the disk is unlikely to
become less than unity in general.  Mass accretion occurs
predominantly along the equatorial plane.  Mass inflow toward the
black hole from the other directions is also found, but mass flow
rates are not large because of smaller density than the value on the
equatorial plane.

We can easily demonstrate that it is not easy to create an extended
magnetosphere which has significantly low-$\beta$ values.  Suppose
that magnetic fields are completely frozen into material.  As a gas
cloud contracts, it gets compressed due to geometrical focusing.  If
the gas cloud shrinks in nearly a spherically symmetric way, the
density and gas pressure increases as $\rho \propto r^{-3}$ and $p
\propto \rho^{\gamma} \propto r^{-5}$ for $\gamma = 5/3$,
respectively, in radiatively inefficient regimes.  Because of the
frozen-in condition, poloidal fields are also amplified according to
the relation, $B_z \propto r^{-2}$, as long as the total magnetic
flux, $B_zr^2$, is conserved.  That is, magnetic pressure also
increases but more slowly, $p_{\rm mag} > {B_z}^2/8\pi \propto
r^{-4}$.  Thus, the plasma-$\beta$ should $increase$ as gas cloud
contracts, as $\beta \equiv p_{\rm gas}/p_{\rm mag} \propto r^{-1}$.
Regions just above and below the black hole (in the vertical
direction) can have low-$\beta$ plasmas, since matter can slide down
to the black hole along the vertical fields, but it is unlikely that
low-$\beta$ regions can occupy a large volume.

If gas cloud shrinks in the two-dimensional fashion, conversely, the
plasma-$\beta$ will decrease as gas cloud contract, since then gas
density is proportional to $\rho \propto r^{-2}$, thus yielding
$p_{\rm gas} \propto r^{-2\gamma} \propto r^{-10/3}$ (for $\gamma =
5/3$) and $\beta \propto r^{2/3}$.  This is not the case in our
simulations, which show that matter is more concentrated around the
equatorial plane.

It might be noted that the surface of a magnetosphere may be unstable
against the magnetic buoyancy instabilities such as interchange
instability (so-called magnetic Rayleigh-Tayer instability; Kruskal \&
Schwarzschild 1954) and undular instability (so-called Parker
instability; Parker 1966), since heavy materials are located above
field lines with respect to the direction of gravity.  Roughly
speaking, the critical wavelength, over which perturbation is unstable
against the Parker instability, is $\lambda_{\rm c} \sim
4\beta^{-1/2}H$ (with $H$ being the pressure scale-height; see
Matsumoto et al. 1988).  When $\beta \sim 10^{-3}$, we have
$\lambda_{\rm c} \sim 100 H > r$, since $H/r$ cannot be as small as
$H/r \sim 10^{-2}$.  The Parker instability is completely stabilized
under such circumstances but interchange instability can occur (e.g.,
Wang \& Nepvue 1983).  In our case with a relatively large $\beta \sim
1$, on the other hand, there are frequent chances to trigger the
Parker instability.  The growth time scale of the Parker instability
is $(2-4)H/v_{\rm A}$. As a result, heavy gas easily sink down through
a moderately magnetized region.

We need to note again that kinetic energy (including orbital energy)
always dominate over magnetic and thermal energy, meaning that there
is no sign of huge magnetic fields inhibiting accretion as INA03
claimed.  It is not clear if magnetic buoyancy instability helps
accreting motion of gas.  Further, we should note that anomalous
resistivity always turns on near the black hole, meaning the effective
occurrence of magnetic reconnection.  In any cases, our picture of the
quasi-steady phase of MHD accretion flow is that matter can accrete
onto a black hole on relatively short timescale.

We studied the case of initially poloidal fields.  A natural question
is how universal our conclusion is.  The simulations starting with
toroidal fields (INA03; Machida \& Matsumoto 2003) show no strong
indication of magnetic jets.  This is a big difference.  However,
toroidal fields can be more easily created by differential rotation
than poloidal fields.  We thus find that toroidal fields can suddenly
grow and dominate over poloidal fields in a large volume (see figure
7).  We can expect similar quasi-steady state, even if we start
calculations with toroidal fields.  To this extent, our study is
consistent with those of initially toroidal field with the low
external pressure.  The comparison is left as future work.

\acknowledgments

Numerical computations were carried out on VPP5000 at the Astronomical
Data Analysis Center of the National Astronomical Observatory, Japan
(yyk27b).   This work was supported in part by the Grants-in Aid of
the Ministry of Education, Science, Sports, Technology, and Culture of
Japan (13640328,14079205, SM).

\end{document}